\documentclass[12pt]{article}
\usepackage{graphicx}
\usepackage{amsmath}

\input{tcilatex}

\begin{document}

\begin{abstract}
The main topics of this paper is to shown a Fast Mie Algorithm FMA as the
best way to use the Mie scattering theory for cross section calculation.
This fast algorithm used recursion for summing a long timed sum of
cylindrical functions.\newpage 
\end{abstract}

\begin{center}
{\Large THE LIGHT SCATTERING AND FAST MIE ALGORITHM}

\vspace{0.2in} Pawel Gliwa {\textit{Sigma Poland\\[0pt]
Warsaw \\[0pt]
}{\small e-mail: \texttt{Pawel.Gliwa@sigmapl.pl}}}
\end{center}

\medskip

\vfill
\newpage

\section{Mie Theory}

\emph{Gustaw Mie} in 1908 had considered a problem of the light scattering
by the macroscopic dielectric spheres. \emph{Mie} find an exact solution of
the problem of diffraction the plane wave on the homogenous sphere with
constant density. Similar solution was published by \emph{Debay}.

The Mie solution satisfies also the scattering on many spheres with the same
structure and diameter -- like scattering on the drops. The \emph{Mie}
solutions in widely used in various situations up to nowadays.

Many scientists (\cite{Gans25}, \cite{Debye09}, \cite{Hulst57}, \cite
{Fabelin} and others) show a lot of simply models, more simply than \emph{%
Mie'}s to application. Unfortunately, all of this approximations, works only
in strictly defined area (sometimes with great error). For precisely
calculation done in all the spectrum of measure we must used the main
theory, now is only one such theory -- \emph{Mie} Theory.

On the long distances, the process of light scattering is described by
amplitude \textbf{A} of electric field \textbf{E}:

\begin{equation}
\mathbf{E}\overset{as}{=}\mathbf{A}\frac{\exp (ikr)}{ikr},\text{ \ }kr>>1.
\end{equation}
The dependencies of angle give us angle distribution like:

\begin{equation}
d\sigma /d\Omega =\left| \mathbf{A}/ik\right| ^{2}.
\end{equation}
After integration over all angles it takes us total cross section. Amplitude
can be write also in such form:

\begin{equation}
\mathbf{A}=\mathbf{n}\times \left( \mathbf{e}\times \mathbf{n}\right) H_{1}+%
\mathbf{n}\times \left( \mathbf{e}\times \mathbf{k}\right) H_{2}
\end{equation}
This form is more useful for calculation of physically dependencies. For
example, amplitude of radiation, with parallel polarisation, is equal to:

\begin{equation}
A_{\Vert }=\left( H_{1}\cos \theta +H_{2}\right) \cos \varphi ,
\end{equation}
and for perpendicular

\begin{equation}
A_{\bot }=-\left( H_{1}+H_{2}\cos \theta \right) \sin \varphi \text{ }.
\end{equation}
For non-polarised wave the intensity of reflected light is equal:

\begin{eqnarray*}
I_{\Downarrow }\left( \theta \right) &=&\frac{1}{2}\left| H_{1}\cos \theta
+H_{2}\right| ^{2}, \\
I_{\bot }\left( \theta \right) &=&\frac{1}{2}\left| H_{1}+H_{2}\cos \theta
\right| ^{2}.
\end{eqnarray*}
These components give us total intensity $I(\theta )$ :

\begin{equation}
I(\theta )=\frac{1+\cos ^{2}\theta }{2}\left( \left| H_{1}\right|
^{2}+\left| H_{2}\right| ^{2}\right) +2\cos \theta \limfunc{Re}\left(
H_{1}H_{2}^{\ast }\right)
\end{equation}
Polarisation $P(\theta )$ is given as:

\begin{equation}
P(\theta )=\sin ^{2}\theta \frac{\left| H_{1}\right| ^{2}-\left|
H_{2}\right| ^{2}}{2I(\theta )}.
\end{equation}
If $H_{2}=0$, then we get the known \emph{Rayleigh}'s results for amplitude

\begin{equation}
\mathbf{A}_{R}=\mathbf{n}\times (\mathbf{e}\times \mathbf{n})H_{1}.
\end{equation}
Intensity is expressed by:

\begin{equation}
I_{R}(\theta \,)=\frac{1+\cos ^{2}\theta }{2}\left| H_{1}\right| ^{2},
\end{equation}
and polarisation as:

\begin{equation}
P_{R}(\theta \,)=\frac{\sin ^{2}\theta }{1+\cos ^{2}\theta }.
\end{equation}

\section{The Mie sum for amplitude H}

We consider electrical field satisfied the Maxwell equations \cite{Gran94}, 
\cite{Gran94Phys}, \cite{Gran95}, \cite{Grn295}. If $t=\cos \vartheta $,
then:

\begin{equation}
H_{1}=(M-tE)\prime ,\quad \;H_{2}=(E-tM)\prime .
\end{equation}
Using an operator notation \cite{Gran95} we can write: 
\begin{eqnarray}
D_{l}^{\left( p\right) } &=&\hat{O}_{p}h_{l}\left( \alpha \right)
j_{l}\left( \beta \right) \\
\hat{O}_{M} &=&\beta \partial _{\beta }-\alpha \partial _{\alpha }  \notag \\
\hat{O}_{E} &=&\alpha \partial _{\beta }-\beta \partial _{\alpha }  \notag \\
\alpha &=&kR  \notag \\
\beta &=&NkR.  \notag
\end{eqnarray}
Both $E$ and $M$ have a well-known structure:

\begin{equation}
E=\sum\limits_{l=1}^{\infty }\frac{2l+1}{l(l+1)}P\prime _{l}(t)E_{l},
\end{equation}

\begin{equation}
M=\sum\limits_{l=1}^{\infty }\frac{2l+1}{l(l+1)}P\prime _{l}(t)M_{l},
\end{equation}
where $P_{l}\left( t\right) $ are the \emph{Legendre} polynomials. Now, $E$
and $M$ can be written as

\begin{equation}
E_{l}=e^{i\delta \,_{l}^{E}}\sin \delta \,_{l}^{E}=\frac{\limfunc{Re}%
D_{l}^{E}}{D_{l}^{E}},
\end{equation}

\begin{equation}
M_{l}=e^{i\delta \,_{l}^{M}}\sin \delta \,_{l}^{M}=\frac{\limfunc{Re}%
D_{l}^{M}}{D_{l}^{M}},
\end{equation}
where $j_{l}(z),\;h_{l}(z)$ are cylindrical functions of Bessel and Hankel
(first kind functions). We can write now the electric part

\begin{equation}
E_{l}=\frac{j_{l}(\alpha \,)j_{l}^{\prime }(\beta \,)-Nj_{l}^{\prime
}(\alpha \,)j_{l}(\beta \,)}{h_{l}(\alpha \,)j_{l}^{\prime }(\beta
\,)-Nh_{l}^{\prime }(\alpha \,)j_{l}(\beta \,)},
\end{equation}
and magnetic one

\begin{equation}
M_{l}=\frac{Nj_{l}(\alpha \,)j_{l}^{\prime }(\beta \,)-j_{l}^{\prime
}(\alpha \,)j_{l}(\beta \,)}{Nh_{l}(\alpha \,)j_{l}^{\prime }(\beta
\,)-h_{l}^{\prime }(\alpha \,)j_{l}(\beta \,)}.
\end{equation}

\section{Approximations}

For small $\alpha $, the scattering cross-section was written by Lord \emph{%
Rayleigh }\cite{Rayleigh11}, \cite{Rayleigh14}, \cite{Rayleigh1871}:

\begin{equation}
Q_{R}=\frac{8\alpha ^{4}}{3}\frac{\left( N^{2}-1\right) ^{2}}{\left(
N^{2}+2\right) ^{2}}.
\end{equation}
For water, this approximation work for $\alpha <0.5$ with error smaller than 
$1\%$. In range of $\alpha >$$>1$, good approximation take us the formula of
van de Hulst \cite{Hulst57}:

\begin{equation}
Q_{H}=2\left[ 1-2\frac{\sin \delta }{\delta }+2\frac{1-\cos \delta }{\delta
^{2}}\right] ,\delta =2\alpha (N-1)\text{ }.
\end{equation}
Scattering cross - section is defined as:

\begin{equation}
\sigma _{sc}=\int d\Omega \left| \mathbf{A}/k\right| ^{2}.
\end{equation}
After normalisation by $\pi a^{2}$ is equal:

\begin{equation}
Q=\frac{2}{\alpha ^{2}}\sum\limits_{l=1}^{\infty }\left( 2l+1\right) \left(
\left| E_{l}\right| ^{2}+\left| M_{l}\right| ^{2}\right) .  \label{Total Q}
\end{equation}
So, we obtained total cross section.

\section{FMA - \emph{Fast Mie Algorithm}}

Our program calculated Eq.$\left( \ref{Total Q}\right) $. This program is
written in Turbo Pascal language. The structure of algorithm is very simple
so the translation to another language can take no more like some minutes.

In program we used recursion, so the time of calculation is minimized to
some seconds for thousand of dates. Recursion used main properties of
calculated functions and is most efficient for great alpha.

Program generates the data file with location C:\TEXTsymbol{\backslash}%
\textsf{MIE.DAT}.

\newpage \textsf{PROGRAM FASTMIE; }$\{$\textsf{simplified source code}$\}$

\textsf{Uses Crt;}

\textsf{Var}

\textsf{Wy:Text;}

\textsf{wave\_k,wave:Integer;}

\textsf{q,n,w,alpha,alpha\_k,alpha\_p,alpha\_step :Extended;}

\textsf{crossection, crossection\_el, crossection\_ml :Extended;}

\textsf{a,b,c,d,e,f, pp, m :Extended;}

\textsf{j0, j1, j2, jb0, jb1, jb2, y0, y1, y2 :Extended;}

\textsf{Begin}

\textsf{ClrScr;}

\textsf{Writeln ('Program FAST MIE');}

\textsf{Writeln;}

\textsf{Writeln ('Give N (water N=3/4)'); Readln(n);}

\textsf{Writeln ('From alpha '); Readln(alpha\_p);}

\textsf{Writeln ('To alpha '); Readln(alpha\_k);}

\textsf{Writeln ('Step of alpha '); Readln(alpha\_step);}

\textsf{Assign (Wy,'C:}$\backslash $\textsf{MIE.DAT');}

\textsf{Rewrite (Wy);}

\textsf{alpha:=alpha\_p;}

\textsf{Repeat}

\textsf{alpha:=alpha+alpha\_step;}

\textsf{crossection:=0;}

\bigskip

$\{$\textsf{\ calculation for 1}$^{\mathsf{st}}$\textsf{\ wave }$\}$

\textsf{wave\_k:=round(alpha*1.12+10);}

$\bigskip $

$\{$\textsf{initial values of cylindrical functions }$\}$

$\{$\textsf{\ j=Re\_h y=Im\_h }$\}$

\textsf{j0:=sin(alpha); y0:=-cos(alpha);}

\textsf{j1:=j0/alpha+y0;y1:=y0/alpha-j0;}

\textsf{jb0:=sin(alpha*n);jb1:=jb0/(alpha*n)-cos(alpha*n);}

\textsf{pp:=1;}

\textsf{a:=y1; b:=y0;}

\textsf{c:=j1; d:=j0;}

\textsf{e:=jb0; f:=jb1;}

\textsf{w:= (n-(1/n))/alpha;}

$\bigskip $

$\{$\textsf{electric part }$\}$

\ \textsf{m:=((a*e)-(n*b*f)+(w*a*f))/((c*e)- (n*d*f)+(w*c*f));}

\ \textsf{q:=(2/(alpha*alpha))* (pp+2);}

\ \textsf{crossection\_el:=q*(1/(1+(m*m)));}

\ \textsf{crossection:=crossection+crossection\_el;}

$\{$\textsf{magnetic part }$\}$

\textsf{m:=((n*a*e)-(b*f))/((n*c*e)-(d*f));}

\ \textsf{crossection\_ml:=q*(1/(1+(m*m)));}

\ \textsf{crossection:=crossection+crossection\_ml;}

$\bigskip $

$\{$\textsf{recursion for all partial waves }$\}$

\textbf{For wave:=2 to wave\_k Do}

\textbf{Begin}

\textsf{pp:=(2*wave-1);}

\textsf{j2:= pp * j1 / alpha - j0;}

\textsf{y2:= pp * y1 / alpha - y0;}

\textsf{j0:=j1; j1:=j2;}

\textsf{y0:=y1; y1:=y2;}

\textsf{a:=y1;b:=y0;}

\textsf{c:=j1;d:=j0;}

\textsf{jb2:=pp * jb1 / (alpha*n) - jb0;}

\textsf{jb0:=jb1; jb1:=jb2;}

\textsf{e:=jb0; f:=jb1;}

\textsf{w:= wave*(n-(1/n))/alpha;}

$\qquad \{$\textsf{electric part }$\}$

\qquad \textsf{m:=((a*e)-(n*b*f)+(w*a*f))/((c*e)-(n*d*f)+(w*c*f));}

\qquad \textsf{q:=(2/(alpha*alpha))* (pp+2);}

\qquad \textsf{crossection\_el:=q*(1/(1+(m*m)));}

\qquad \textsf{crossection:=crossection+crossection\_el;}

$\qquad \{$\textsf{\ magnetic part }$\}$

\qquad \textsf{m:=((n*a*e)-(b*f))/((n*c*e)-(d*f));}

\qquad \textsf{crossection\_ml:=q*(1/(1+(m*m)));}

\qquad \textsf{crossection:=crossection+crossection\_ml;}

\textbf{End;}

\textsf{Writeln(Wy, alpha:5:5, ' ' , crossection:5:5);}

\textsf{Until (alpha}$>$\textsf{=alpha\_k);}

\textsf{Close(Wy);}

\textsf{End.}\newpage

\section{Results}

Now, we can show some results as plots for $N$ equal to 4/3, 1.1. On all
these plots we can see very interesting structure with resonances Fig 3,4
and 5.

\bigskip 

Fig. 1 The cross - section for N=1.1 for alpha from 0 to 100.

Fig. 2 The cross --section for N=4/3 for alpha from 0 to 100.

Fig. 3 The cross - section for N=4/3 for alpha from 60 to 65.

Fig. 4 The cross-section for N=4/3 for alpha from 63 to 64.

Fig. 5 The cross-section for N=4/3 for alpha from 63.30 to 63.34. We can see
the Lorenz curve.

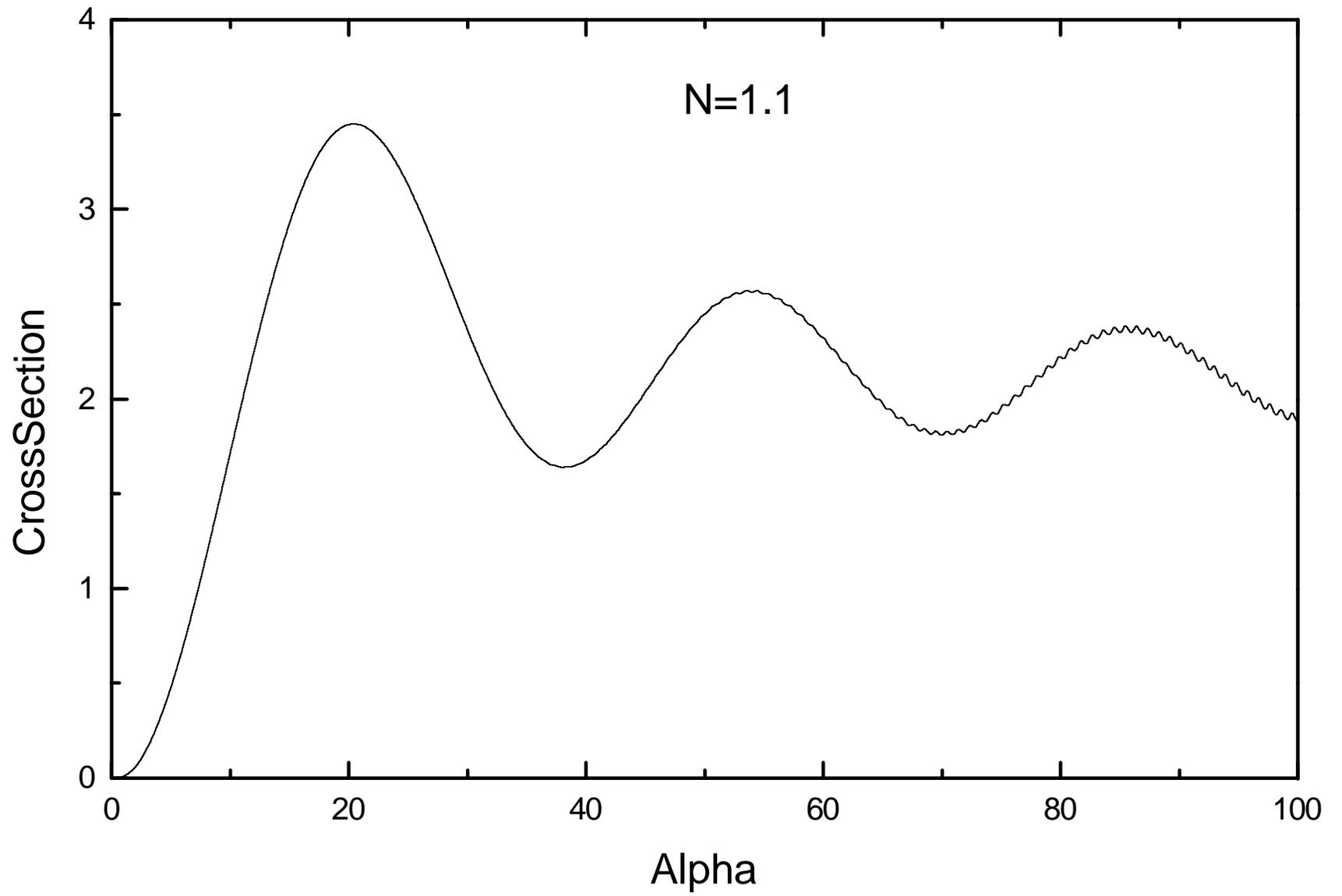
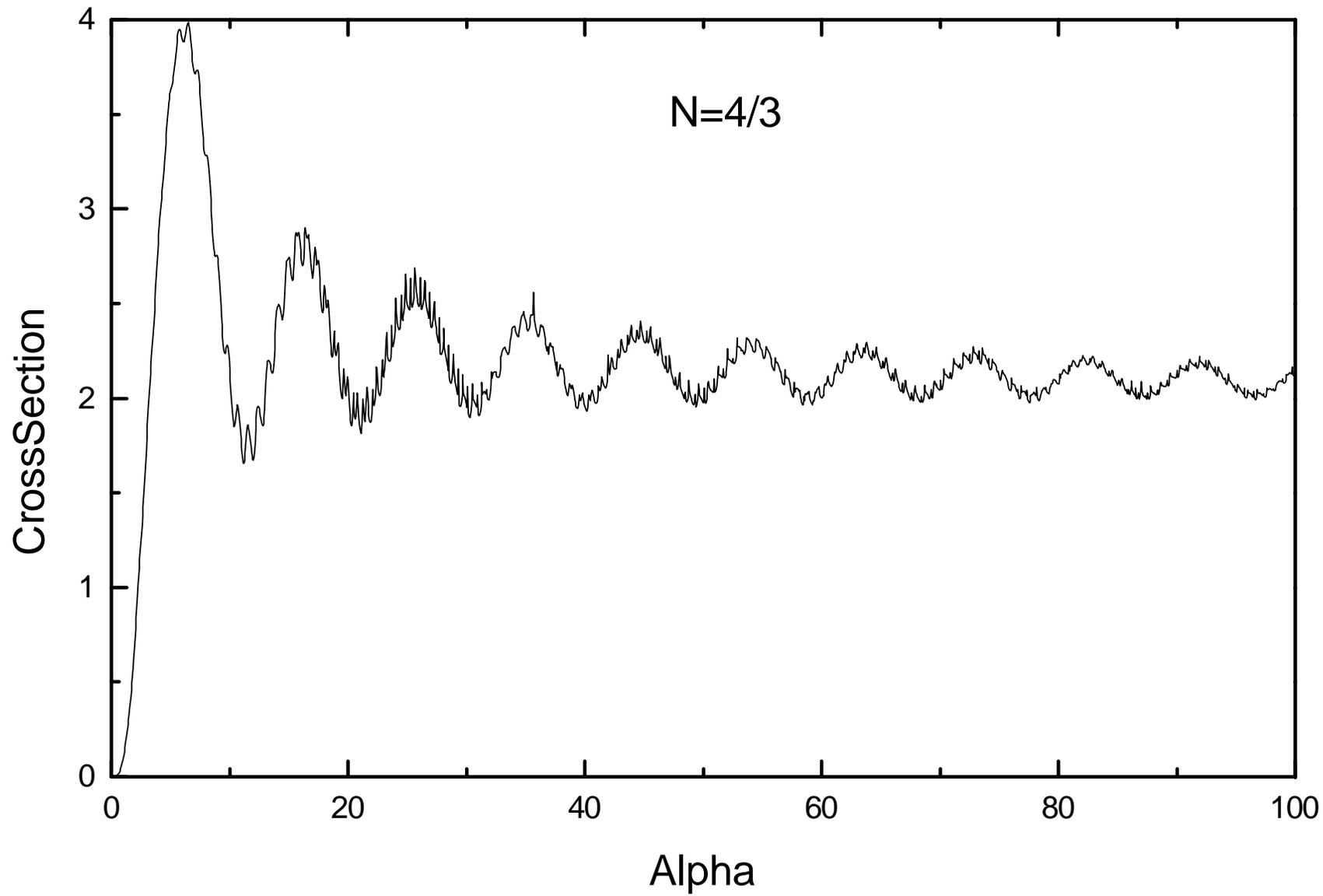
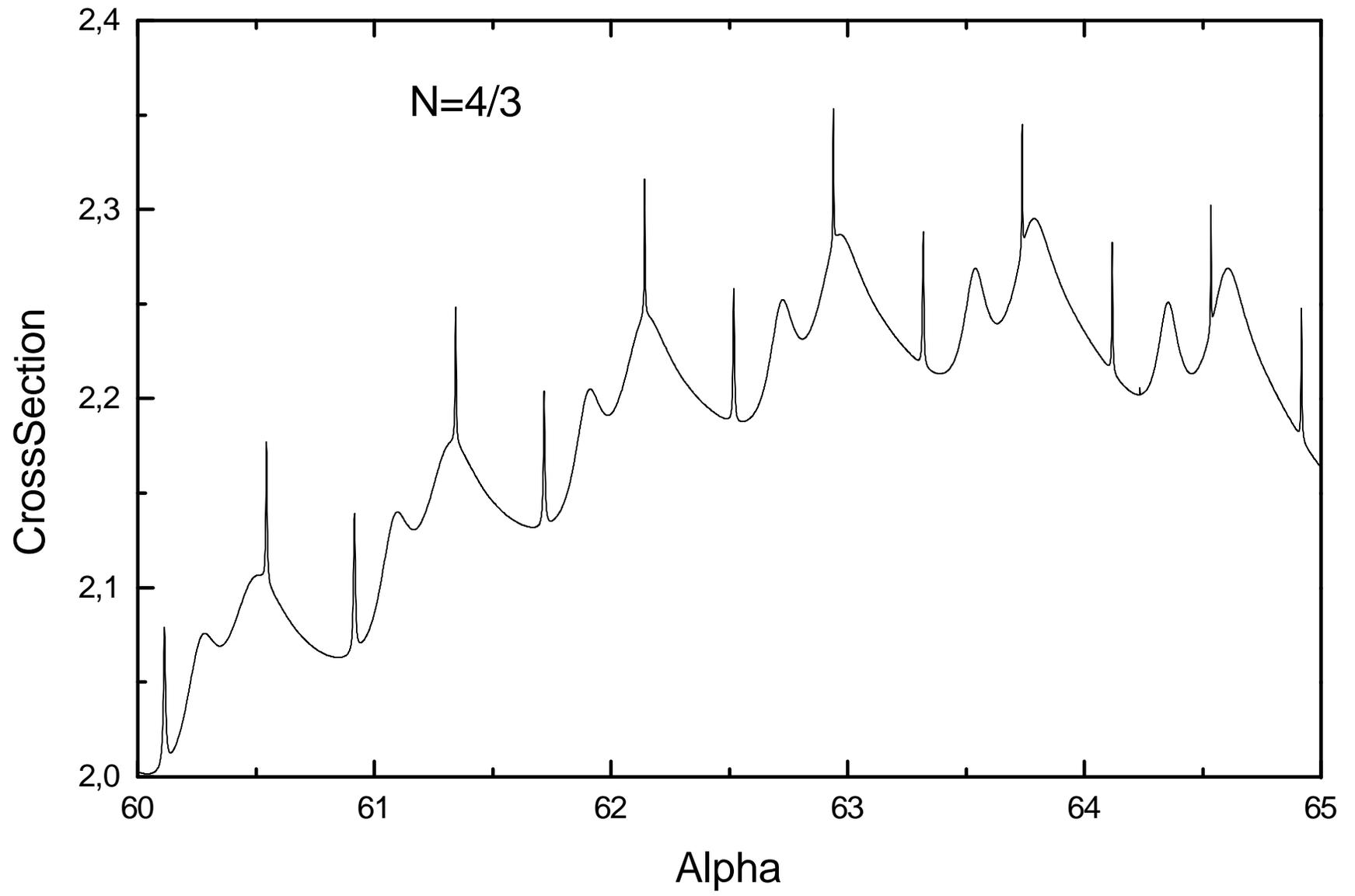
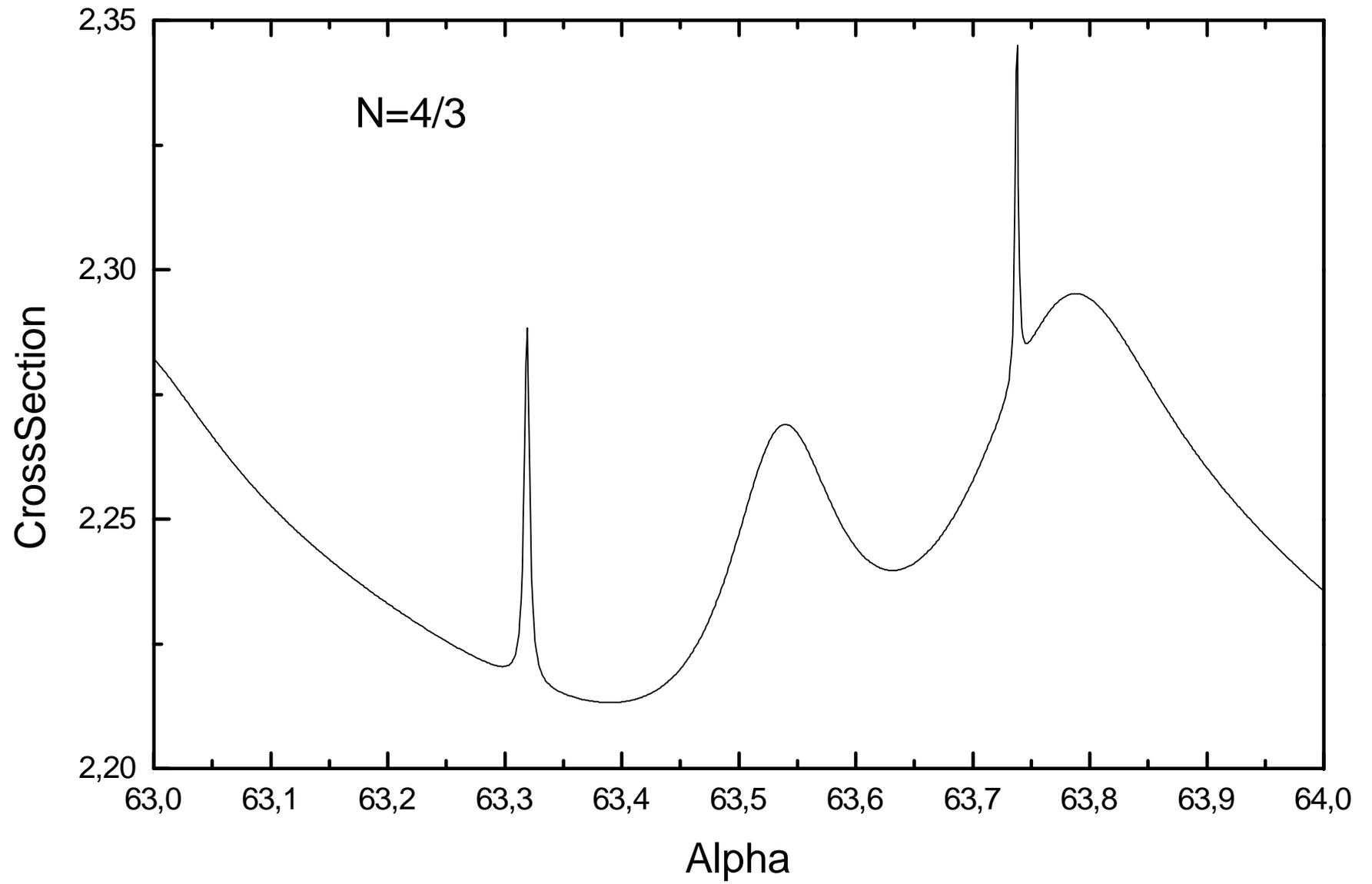
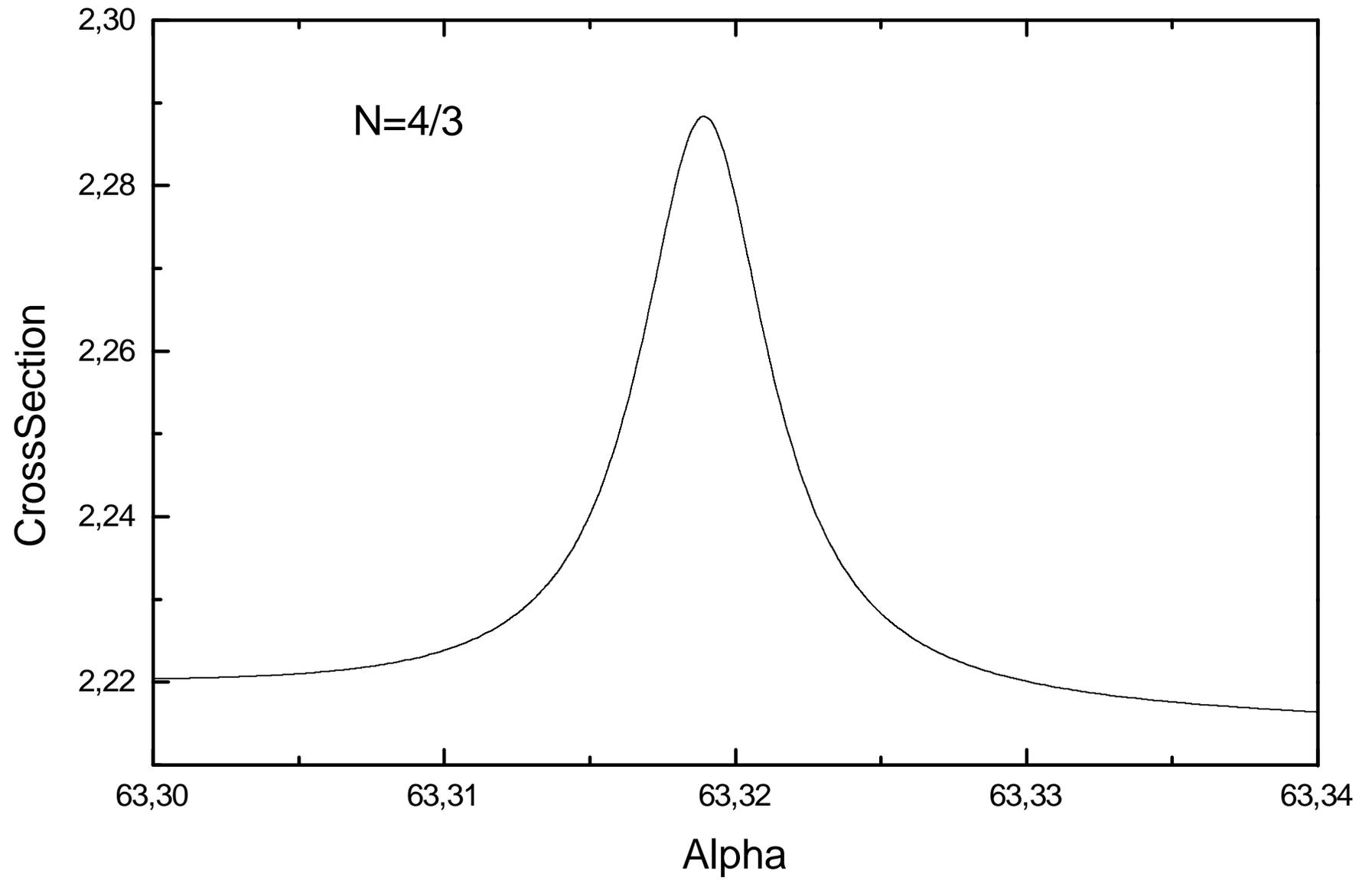
\end{document}